\documentclass{pasj00}
\draft

\begin{document}
\SetRunningHead{Kohno et al.}{GRB 030329: Molecular Gas in the Host Galaxy}
\Received{2004/03/31}
\Accepted{2004/10/30}

\title{
Nobeyama Millimeter Array Observations of GRB 030329:
a Decay of Afterglow with Bumps
and Molecular Gas in the Host Galaxy
}

%
 \author{%
   Kotaro \textsc{Kohno},\altaffilmark{1}
   Tomoka \textsc{Tosaki},\altaffilmark{2}
   Takeshi \textsc{Okuda},\altaffilmark{3,1}
   Kohichiro \textsc{Nakanishi},\altaffilmark{3}\\ 
   Takeshi \textsc{Kamazaki},\altaffilmark{3}
   Kazuyuki \textsc{Muraoka},\altaffilmark{1}
   Sachiko \textsc{Onodera},\altaffilmark{1}\\
   Yoshiaki \textsc{Sofue},\altaffilmark{1}
   Sachiko K. \textsc{Okumura},\altaffilmark{3}
   Nario \textsc{Kuno},\altaffilmark{3}
   Naomasa \textsc{Nakai},\altaffilmark{3}\thanks{Present address: Graduate Scool of Pure and Applied Science, University of Tsukuba, 1-1-1 Ten-noudai, Tsukuba, Ibaragi 305-8571} \\
   Kouji \textsc{Ohta},\altaffilmark{4}
   Sumio \textsc{Ishizuki},\altaffilmark{5}
   Ryohei \textsc{Kawabe},\altaffilmark{5}
   and    
   Nobuyuki \textsc{Kawai}\altaffilmark{6,7}
	 }

 \altaffiltext{1}{Institute of Astronomy, The University of Tokyo, 
 2-21-1 Osawa, Mitaka, Tokyo 181-0015}
 \email{kkohno@ioa.s.u-tokyo.ac.jp}
 \altaffiltext{2}{Gunma Astronomical Observatory, Nakayama, Takayama, 
                  Agatsuma,Gunma 377-0702}
 \altaffiltext{3}{Nobeyama Radio Observatory, Minamimaki, Minamisaku, Nagano 384-1805}
 \altaffiltext{4}{Department of Astronomy, Kyoto University, Kyoto 606-8502}
 \altaffiltext{5}{National Astronomical Observatory of Japan, 2-21-1 Osawa, Mitaka, Tokyo 181-8588}				  
 \altaffiltext{6}{Department of Physics, Tokyo Institute of Technology 2-12-1
 Ookayama, Meguro-ku, Tokyo 152-0033}
 \altaffiltext{7}{Cosmic Radiation Laboratory, The Institute of Physical and Chemical Research (RIKEN),\\
 2-1 Hirosawa, Wako, Saitama 351-0198}

\KeyWords{galaxies: ISM---gamma rays: bursts---gamma rays: individual (GRB 030329)} 

\maketitle

\begin{abstract}
Nobeyama Millimeter Array was used to observe millimeter-wave afterglow of 
GRB 030329 at 93 GHz and 141 GHz 
from 2003 April 6 (8.23 days after the burst)
to 2003 May 30 (61.97 days).
A sensitive search for CO $(J=1-0)$ emission/absorption from the host galaxy
of GRB 030329 was also carried out.
Unresolved millimeter continuum emission at the position of GRB 030329
was detected until 2003 April 21.
We found a steep decline of continuum flux ($\propto t^{-2.0}$) 
during this period,
in accord with a previous report.
Moreover, our data implies that the decay was accompanied 
by possible plateaus phases, or bumps, on a time scale of several days.
From an integrated spectrum, produced by summing up the data 
from 2003 April 10 to 2003 May 30,
we found a possible emission feature, 
which could be a redshifted CO $(J=1-0)$ line.
Its position and redshift coincide well with those of GRB 030329,
though further observations are required to confirm the detection.
If the emission feature is real,
the observed CO flux is $1.4 \pm 0.52$ Jy km s$^{-1}$,
corresponding to a large molecular gas mass of
$M$(H$_2$) $> 10^9$ $M_\odot$.
This implies that the host galaxy, which is optically faint, 
is highly obscured due to a rich interstellar medium.
\end{abstract}

\section{Introduction}

It is now widely believed that long-duration gamma-ray bursts (GRBs) are 
extremely energetic phenomena associated with star formation 
in galaxies at a cosmological distance. 
Detection of the supernova component SN 2003dh (Stanek et al.\ 2003; 
Hjorth et al.\ 2003; Kawabata et al.\ 2003) in the afterglow of the HETE-2
localized GRB 030329 (Vanderspek et al.\ 2004) has established 
the association of some of the GRBs 
with core-collapse supernovae from massive stars.
GRBs are often hosted by dusty, 
star-forming galaxies with elevated star formation rates (SFRs) 
of a few -- 100 $M_\odot$ yr$^{-1}$
(Berger et al.\ 2001, 2003a; Sokolov et al.\ 2001; 
Frail et al.\ 2002; Chary et al.\ 2002). 
To sustain such putative vigorous star formation, 
it is immediately clear that a large amount of molecular gas, 
a fuel of massive star formation, must exist
in the host galaxies of GRBs. 
For example, a molecular gas mass of $\sim 10^{9} M_\odot$ 
is required in order to keep a SFR of 100 $M_\odot$ y$^{-1}$ for $10^7$ yr, 
which is a typical duration of starburst 
(e.g., Thornley et al.\ 2000, and references therein).
It is also claimed that GRBs with undetected, or dark, optical afterglows (``dark bursts'')
probably occur in molecular clouds (Reichart, Price 2002; but see Barnard et al.\ 2003); 
if it is the case, the surrounding interstellar medium (ISM) would be drastically affected 
by GRBs (Draine, Hao 2002). In fact, high column densities 
up to $N_{\rm H} \sim 10^{22 - 23}$ cm$^{-2}$,
consistent with the idea of GRBs within molecular clouds, are suggested 
based on an analysis of GRB afterglows at X-ray (Galama, Wijers 2001).

Nevertheless, no observations of molecular gas have been reported toward GRB hosts, so far.
This is basically due to the limited sensitivities of the current millimeter-wave facilities.
Many of GRBs are at a cosmological distance ($z>1$), but CO emission from galaxies beyond $z=1$
is not very easy to observe, unless it is gravitationally amplified or extremely luminous
(e.g., Barvainis et al.\ 2002).

Here, we present a sensitive search for redshifted CO $(J=1-0)$ emission 
from the host galaxy of the luminous burst GRB 030329/SN 2003dh
using the Nobeyama Millimeter Array (NMA).
The continuum emission at millimeter wavelengths was also monitored simultaneously.
GRB 030329 is one of the closest objects found among the luminous GRBs to date 
($z$ = 0.1685; Greiner et al.\ 2003; Caldwell et al.\ 2003),
and therefore a suitable target for a deep search of molecular gas.

Millimeter-wave continuum emission from afterglow is also of great interest.
Because the spectral energy distribution (SED) 
of afterglow peaks at millimeter wavelengths 
on a timescale of a few days after a burst
(e.g., Berger et al.\ 2000),
flux monitoring at mm-wavelengths is very important 
to determine the peak flux and frequency of the synchrotron spectrum 
as a function of time from the burst (e.g., Galama et al.\ 2000).
Flux monitoring of afterglow at millimeter wavelengths has been made 
toward several bright GRBs,
such as GRB 970508, GRB 980329, GRB 991208, GRB 000301C, GRB 010222, 
and GRB 030329 
(e.g., Bremer et al.\ 1998; Taylor et al.\ 1998; 
Galama et al.\ 2000; Berger et al.\ 2000; Frail et al.\ 2002; Galama et al.\ 2003; 
Sheth et al.\ 2003; Kuno et al.\ 2004).

Assuming a cosmology with $H_0$ = 71 km s$^{-1}$ Mpc$^{-1}$,
$\Omega_{\rm M}$ = 0.27, and $\Omega_{\Lambda}$ = 0.73,
the luminosity distance of GRB 030329 is $d_{\rm L}$ = 805 Mpc,
and the angular distance is $d_{\rm A}$ = 589 Mpc ($1''$ corresponds to 2.86 kpc).

\section{Observations and Data Analysis}

We observed GRB 030329 using Nobeyama Millimeter Array (NMA) 
during the period from 2003 April 6 to 2003 May 30.
NMA consists of six 10 m dishes equipped with cryogenically cooled receivers 
employing Superconductor-Insulator-Superconductor (SIS) mixers in double side band (DSB) operation.
The most compact configuration (D array; baseline lengths ranged from 13 m to 82 m) was used 
for the observations, except for the first two days
(C array for April 6 and 7; baseline lengths were from 26 m to 163 m). 
For 3 mm observations, the tracking frequency was set to 98.825 GHz 
to observe a redshifted CO(1--0) line ($\sim$ 98.65 GHz) at the upper side band (USB), 
whereas the lower side band (LSB), centered at 86.825 GHz, 
(and line-free channels of USB) was for the continuum.
The center frequencies of USB and LSB for 2 mm observations were 
146.969 GHz and 134.969 GHz, respectively.
Each side band was separated by 90$^\circ$ phase switching of a reference signal.
The Ultra Wide-Band Correlator (Okumura et al.\ 2000) was configured 
to cover a bandwidth of 1024 MHz per side-band with an 8 MHz resolution.

A radio source, J1159+292, was observed every $\sim$ 20 minutes 
for amplitude and phase calibrations,
and the passband shape of the system was determined 
from observations of strong continuum sources, 3C 273 or J0423--013.
We also observed J1058+015 several times during each observing run 
to check the consistency of the amplitude calibration. 
The flux density of J1159+292 was determined several times during the observing run;
it ranged from 1.6 to 2.3 Jy. In this paper, we adopted a constant flux value of 2.0 Jy 
throughout our observing run. This gave almost the same flux scale 
as the OVRO/BIMA results (Sheth et al.\ 2003; 1.95 Jy was adopted to the data).
The overall error of the absolute flux scale was estimated to about $\pm$ 20\%.

The raw visibilities were edited and calibrated using the NRO UVPROC-II package 
(Tsutsumi et al.\ 1997),
and then USB and LSB data were concatenated to produce a final continuum image
of each day using the NRAO AIPS task IMAGR.

For a spectral line analysis, the continuum emission was subtracted first 
from the visibilities using the UVPROC-II task LCONT, if continuum was detected.
Then, each data set was combined into a single visibility set, 
and channel maps with a width of 32 MHz (97.2 km s$^{-1}$)
at an increment of 16 MHz (48.6 km s$^{-1}$) were produced.

\section{Results}

\subsection{Continuum Emission}

An unresolved 2 mm/3 mm continuum source was detected at the position of GRB 030329 
from 2003 April 6 to April 21. 
The observational log is provided in table 1.
Some selected continuum images, showing a fading afterglow emission,
are presented in figure 1,
and the obtained millimeter-wave light curve is given in figure 2.

The visibilities often suffered from large phase noises of atmospheric origin.
To estimate the effect of the phase noise, 
we made images of the visibility calibrator source J1159+292 as well. 
We found that the decorrelation factors were typically $\sim$ 0.95,
which decreased to 0.90 for the worst case. 
The obtained flux densities listed in table 1
are after a correction of the decorrelation.

\subsection{Search for Spectral Line Emission/Absorption}

Neither significant emission nor an absorption feature was detected 
from individual spectra of each day.
This is consistent with a report by Sheth et al.\ (2003).
However, by summing up the data at 3 mm from 2003 April 10 to May 30,
we found a possible emission feature from this integrated spectrum.
Figure 3 displays the spectrum, 
and a velocity-integrated image of the possible emission feature.
Note that the data on April 15 was not included in the final spectrum,
because the weight of the data was very small due to its larger system noise temperature.

The spectrum shows an emission feature with a peak flux density of $6.2 \pm 3.8$ mJy
at 98.62 GHz, corresponding to a redshift of 
$z_{\rm CO} = 0.1689$, if this is CO $(J=1-0)$ emission.
The significance of the emission feature is low ($\sim 2 \sigma$), 
but this feature appears on several successive channels,
from $z_{\rm CO} = 0.1682$ to 0.1689.
This is within the range of the reported GRB 030329 redshifts determined
from optical emission/absorption lines,
0.1685 (Greiner et al.\ 2003) to 0.169 $\pm$ 0.001 (Price et al.\ 2003).

The velocity width of this feature is $220 \pm 49$ km s$^{-1}$.
If we integrated the possible emission feature over the velocity width of 220 km s$^{-1}$,
we found a $\sim 3 \sigma$ source at the position of GRB 030329 (figure 3, right panel).
The integrated CO flux density is $1.4 \pm 0.52$ Jy km s$^{-1}$.
The position and redshift coincide well with those of GRB 030329, 
though we need further observations
to confirm this possible emission-line feature.

\section{Discussion}

\subsection{Millimeter-Wave Light Curve: a Rapid Decay with Possible Plateaus?}

The overall appearance of our millimeter-wave light curve in table 1 and figure 2
is consistent with the OVRO/BIMA (Sheth et al.\ 2003) 
and NRO 45 m telescope (Kuno et al.\ 2004) results;
i.e., a rapid decay of millimeter-wave afterglow has been observed.
Figure 4 shows a comparison of $\lambda$ = 3 mm measurements of GRB 030329 afterglow
from NMA, OVRO/BIMA, and the 45 m telescope.
A least-squares fit of all 3 mm observations 
during $\Delta t$ = 9 to 23 gives
a temporal decay index of $\alpha_{\rm 3 mm} = -2.03 \pm 0.21$, 
where millimeter flux, $S$, decays as $S \propto t^\alpha$.
The steep slope observed at millimeter-wavelengths, 
close to the decay indices 
after the jet break at $\Delta t \sim 0.5$ days
at optical wavelengths ($\alpha = -1.97 \pm 0.12$, Price et al.\ 2003; 
$\alpha = -1.81 \pm 0.04$, Sato et al.\ 2003),
suggests that the expansion during this period ($\Delta t$ = 9 to 23) 
still remained relativistic (e.g., Frail et al.\ 2004).

In addition to the rapid decay,
our data also implies that the decay was accompanied by 
possible ``plateau'' phases, or ``bumps'',
with a time scale of $\sim$ several days.
One can see that the flux densities in figure 2
are almost constant from $\Delta t$
= 12 to 17 days (three days average is 17.6 mJy), 
whereas it drops suddenly at $\Delta t$ = 18, and then 
the light curve becomes constant again till $\Delta t$ = 23 (average is 9.0 mJy).
The difference between these two plateau levels, a factor of 2, is significantly larger 
than the estimated overall flux uncertainty, $\sim$ 20\%.
In figure 4, these possible plateaus/bumps are not very evident, however. 
This could be because the fluxes in figure 4 were determined from
different instruments, causing a significant error in the absolute calibration.

Note that the detection of fluctuation at millimeter-wavelengths is important
because the effect of interstellar scattering/scintillation, 
which could be significant for radio light curves
(e.g., Frail et al.\ 2000), is mitigated at millimeter-wave light curves
(\cite{walker1998}).

It is intriguing to see whether these possible bumps on the millimeter-wave light curve 
are achromatic or not,
because simultaneous multi-wavelengths light curves provide a crucial clue 
to the origin of the bumps on the light curves (e.g., Nakar et al.\ 2003).
Unfortunatelly, a detailed optical light curve compiled by
Lipkin et al.\ (2004) lacks data points from $\Delta t$ $\sim$ 14 to 21 days,
where we find millimeter-wave bumps. Bloom et al.(2004) also presents late-phase
($\Delta t > 10$ days) optical/near-infrared light curves, yet their light curves
also have gaps from $\Delta t$ = 13 to 23 days.

Some of the bumps or rebrightening events seen in optical/NIR bands are often explained 
by a ``refreshed shock'' model (e.g., Granot et al.\ 2003),
yet it remains unclear whether radio/millimeter-wave light curves can also be understood
by this model (Berger et al.\ 2003b; Sheth et al.\ 2003; Frail et al.\ 2004).
If the observed plateaus/bumps in the millimeter-wave light curve
are caused by density fluctuations of surrounding ISM 
(e.g., Wang, Loeb 2000), 
the spatial scale of the structure corresponding to the time scale of the observed bumps, 
several days, is about $\sim 10^3$ AU, supposing a relativistic expansion.
Small-scale structures of a dense molecular medium with such a spatial scale, 
a few 100 AU to 1000 AU, have been found in the Galactic star-forming regions 
(Kamazaki et al.\ 2001).

\subsection{CO Line from the Host Galaxy: a Very Gas/Dust Rich, Optically Faint Galaxy?}

Although the detection of the CO line from GRB 030329 should be confirmed
by additional/independent observations,
if the emission-line feature in figure 3 is real, 
it gives valuable information about ISM 
in the host galaxy of GRB 030329.

\subsubsection{Molecular gas mass}

The integrated CO flux, $1.4 \pm 0.52$ Jy km s$^{-1}$, gives 
a coarse estimation of the molecular gas mass 
using a formula (Solomon et al.\ 1992), as follows:
\begin{eqnarray*}
M({\rm H_2}) & = & \alpha_{\rm CO} \cdot L'_{\rm CO} = \alpha_{\rm CO} \cdot \left( \frac{c^2}{2k} \right) 
                 S_{\rm CO}\Delta v \cdot
                 d_{\rm L}^2 \cdot
                 \nu_{\rm rest}^{-2}
                 \left( 1+z \right)^{-1} \\
             & = & 5.5 \times 10^9 
                   \left[ \frac{\alpha_{\rm CO}}{2.89 \mbox{\ $M_\odot$ (K km s$^{-1}$ pc$^2$)$^{-1}$} } \right]  
                   \left( \frac{S_{\rm CO}\Delta v}{1.4 \mbox{\ Jy km s$^{-1}$}} \right) \\
             &   & \cdot 
                   \left( \frac{d_{\rm L}}{805 \mbox{\ Mpc}} \right)^2
                   \left( 1+\frac{z}{0.1685} \right)^{-1} 
                   \mbox{\ } M_\odot,
\end{eqnarray*}
where $\alpha_{\rm CO}$ is the CO luminosity to H$_2$ molecular gas mass conversion factor,
$L'_{\rm CO}$ is the CO line luminosity, 
$S_{\rm CO} \Delta v$ is the velocity-integrated CO flux, and
$d_{\rm L}$ is the luminosity distance.
The CO line luminosity, $L'_{\rm CO}$, is $1.9 \times 10^9$ [K km s$^{-1}$ pc$^2$].
The deduced mass reaches $5.5 \times 10^9$ $M_\odot$, 
if a Galactic conversion factor, 
$X_{\rm CO} = 1.8 \times 10^{20}$ [cm$^{-2}$ (K km s$^{-1}$)$^{-1}$] 
(Dame et al.\ 2001) or 
$\alpha_{\rm CO} = 2.89$ in a unit of [$M_\odot$ (K km s$^{-1}$ pc$^2$)$^{-1}$], is applied.

The molecular gas mass derived here is comparable to 
those in massive and luminous spiral galaxies 
(e.g., Nishiyama, Nakai 2001, and references therein),
and significantly larger than the molecular gas masses in metal-poor dwarf galaxies. 
For instance, the total molecular gas mass of the Large Magellanic Cloud (LMC) is 
$M$(H$_2$) = $5 \times 10^8 M_\odot$ (Fukui et al.\ 1999), 
adopting an $X_{\rm CO}$ of $9 \times 10^{20}$ [cm$^{-2}$ (K km s$^{-1}$)$^{-1}$],
determined from their own observations.
Similarly, the molecular gas mass of the Small Magellanic Cloud (SMC) is 
$M$(H$_2$) = $4 \times 10^6 M_\odot$ 
by applying $X_{\rm CO} = 2.5 \times 10^{21}$ [cm$^{-2}$ (K km s$^{-1}$)$^{-1}$]
(Mizuno et al.\ 2001),
though this is a lower limit 
because their observation covered only a part of the SMC.

Accordingly, the suggested molecular gas mass up to $10^9 M_\odot$ in GRB 030329
is surprisingly large if one recalls that 
the host of GRB 030329 is rather faint in optical bands; 
the absolute magnitude of the host galaxy is just comparable to that of SMC
(Fruchter et al.\ 2003).

One of the error sources of the deduced molecular gas mass is 
the CO luminosity to H$_2$ molecular gas mass conversion factor.
The molecular gas mass becomes even larger (by a factor of $\sim$ 10) 
if we apply an $X_{\rm CO}$ valid for low-metal galaxies (e.g., Arimoto et al.\ 1996).
On the other hand, $X_{\rm CO}$ factors derived in ultra-luminous infrared galaxies (ULIRGs) 
can reduce the gas mass by a factor of $\sim$ 3 -- 4 (Downes, Solomon 1998).
In any case, supposing that the detection is real, 
a large amount of molecular gas, exceeding $10^9 M_\odot$, must be present 
in the host galaxy of GRB 030329.

\subsubsection{Dynamical mass}

If the velocity width of the possible CO emission feature originated 
from rotation of the molecular gas,
one can estimate an enclosed mass in the central region of the host galaxy 
by assuming the radius of a rotating gas disk and its inclination angle.
The enclosed mass, often referred to as the dynamical mass $M_{\rm dyn}$, is 
\begin{eqnarray}
M_{\rm dyn} & = & \frac{r \cdot V_{\rm rot}^2(r)}{G} 
              =   2.3 \times 10^5 \left( \frac{r}{\mbox{kpc}} \right)
                                \left[ \frac{V_{\rm rot}(r)}{\mbox{km s$^{-1}$}} \right]^2 \mbox{\ } M_\odot \\ 
            & = & 3.2 \times 10^{10} \left( \frac{r}{5.7 \mbox{\ kpc}} \right) 
                                 \left( \frac{\Delta V/2}{110 \mbox{\ km s$^{-1}$}} \right)^2
                                 \left[ \sin \left( \frac{i}{45} \right) \right]^{-2} \mbox{\ } M_\odot,
\end{eqnarray}
where $r$ is the radius of the gas disk, 
$V_{\rm rot}(r)$ is an on-plane rotation velocity at a radius of $r$,
$\Delta V$ is an observed line width, 
and $i$ is the inclination of the disk ($i=0^\circ$ is face-on).
Here, we adopt the radius of the disk as $r$ = \timeform{2"} (or 5.7 kpc), 
an upper limit on the source size from our CO map, 
and an inclination angle of $45^\circ$, which is difficult to determine at this moment.
In this case, the molecular gas mass shears about 17\% of the total (dynamical) mass 
in the central kiloparsecs of the host galaxy.

The major uncertainty of $M_{\rm dyn}$ is inclination. 
If the disk is close to edge-on ($i = 80^\circ$ for instance),
$M_{\rm dyn}$ is reduced to be $1.6 \times 10^{10} M_\odot$. 
On the other hand, $M_{\rm dyn}$ could exceed $10^{11} M_\odot$
when the disk is closer to face-on ($i<23$). 
$M_{\rm gas}/M_{\rm dyn}$ is then $\sim$ 1\% to 10\%, 
which is a typical range seen in nearby gas-rich spiral galaxies
(e.g., Nishiyama, Nakai 2001).

\subsubsection{Molecular gas and star formation in the host of GRB 030329}

A gas-rich condition in the host galaxy of GRB 030329, suggested by our observations, 
implies the presence of an intense star-formation/starburst there
because molecular gas masses roughly correlate with the rates of massive star formation
(e.g., Young et al.\ 1996).
However, a SFR in the host of GRB 030329,
determined from an extinction-corrected H$\alpha$ luminosity, is
0.5 $M_\odot$ yrs$^{-1}$ (Matheson et al.\ 2003),
which is quite modest and not so enhanced compared with starburst galaxies
(e.g., Kennicutt 1998). 
Here, we should note that
the SFRs derived from the optical emission lines could underestimate
the true rate of massive star formation if it is deeply embedded within
dense molecular material.
In fact, 
it is reported that the SFRs in some GRB hosts 
(such as GRB 980703, GRB 000210, GRB 000418, and GRB 010222),
derived from optical observations, 
are much lower (by an order of magnitude) 
than those from radio/submillimeter measurements, 
even after an  extinction correction 
(Berger et al.\ 2001, 2003a; Frail et al.\ 2002; Gorosabel et al.\ 2003). 
This situation suggests that optical emission does not fully trace 
the whole massive star-forming regions
in these GRB hosts; this could be the case in GRB 030329.

For instance, it is likely that the unobscured SFR in the host galaxy of GRB 030329 is 
10-times larger than the SFR derived from optical data,
as in the case of many GRB hosts (Berger et al.\ 2003). 
Then, the ``true'' SFR could be $\sim$ 5 $M_\odot$ yrs$^{-1}$, 
and the expected radio (1.4 GHz) and submillimeter (350 GHz) fluxes, 
originated from star formation in the host galaxy, 
would be $\sim$ 0.3 mJy and $\sim$ 0.1 mJy, respectively, 
according to fomulae by Carilli and Yun (1999). 
Such a flux at 1.4 GHz should be detectable
using VLA when the afterglow emission has been disapperaed.

Consequently, further multiwavelengths studies from radio to infrared, 
as well as additional deeper CO observations,
are indispensable to unveil the true properties of star formation and
to confirm the presence of a large amount of ISM, in the host galaxy of GRB 030329.

\vspace{0.5cm}

We would like to acknowledge the referee, D.A.\ Frail, for his invaluable comments.
We thank to NRO and all people who enabled us to carry out 
this target-of-opportunity program.
We are grateful to the NRO/NMA staff for the operation and 
for continuous efforts to improve the NMA.
We extend our gratitude to E.\ Nishihara for notification of GRB 030329,
and to A.\ Endo for careful reading of the manuscript.
KK was financially supported by The Mitsubushi Foundation, 
JSPS Grant-in-Aid for Scientific Research (B) No.14403001 and 
MEXT Grant-in-Aid for Scientific Research on Priority Areas No.15071202. 


\begin{figure}
  \begin{center}
    \FigureFile(150mm,40mm){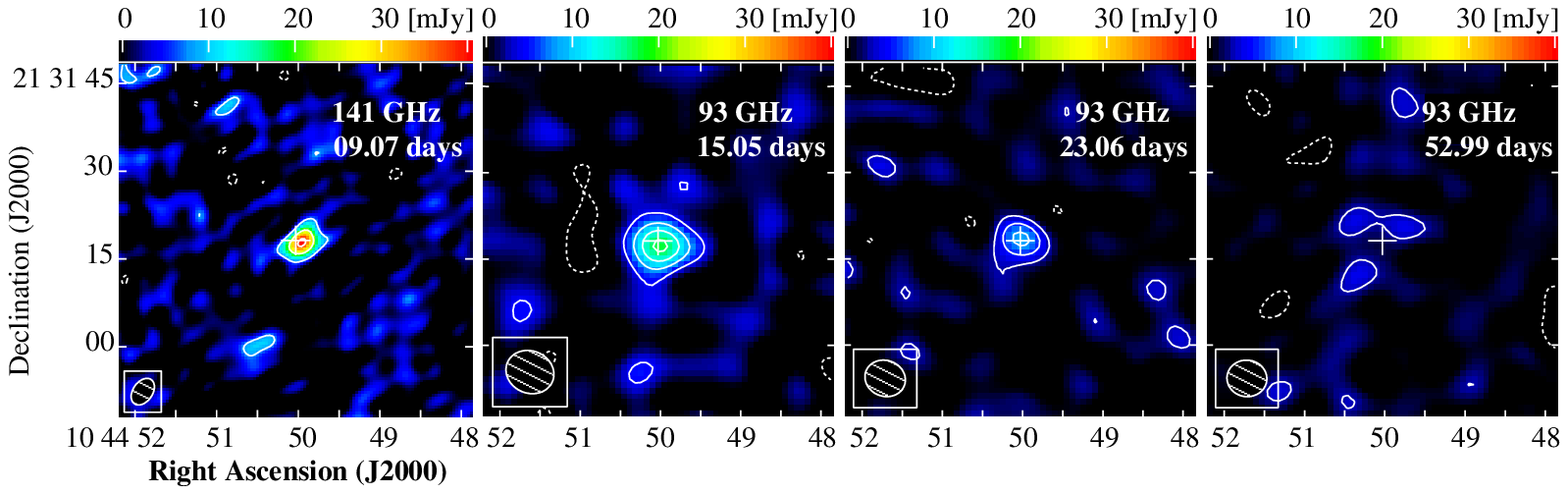}
  \end{center}
\caption{
Selected images of GRB 030329 afterglow at $\lambda$ = 2 mm and 3 mm
to depict fading afterglow emission. 
The observed frequency, time after the burst, and the synthesized beam are indicated.
The central cross in each panel marks the position of the GRB 030329 optical afterglow,
$\alpha_{\rm J2000}$ = \timeform{10h44m50s.03} and 
$\delta_{\rm J2000}$ = \timeform{+21d31'18".15} (Uemura et al.\ 2003).
The contour levels are $-3$ $\sigma$, 3 $\sigma$, 9 $\sigma$, and 15 $\sigma$
for the 141 GHz image, and the contour intervals are 2 $\sigma$ for 93 GHz images.
The $\sigma$ of each day is listed in table 1.
}\label{fig:contmaps}
\end{figure}

\begin{figure}
  \begin{center}
    \FigureFile(80mm,80mm){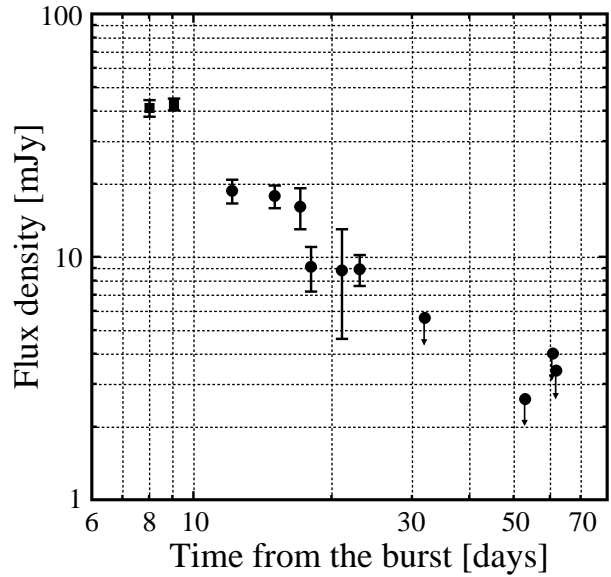}
  \end{center}
\caption{
Time variation of $\lambda$ = 2 mm (square) and 3 mm (circle) continuum flux
toward GRB 030329 observed with the NMA. The upper limits are 2 $\sigma$.
}\label{fig:lightcurve}
\end{figure}

\begin{figure}
  \begin{center}
    \FigureFile(150mm,80mm){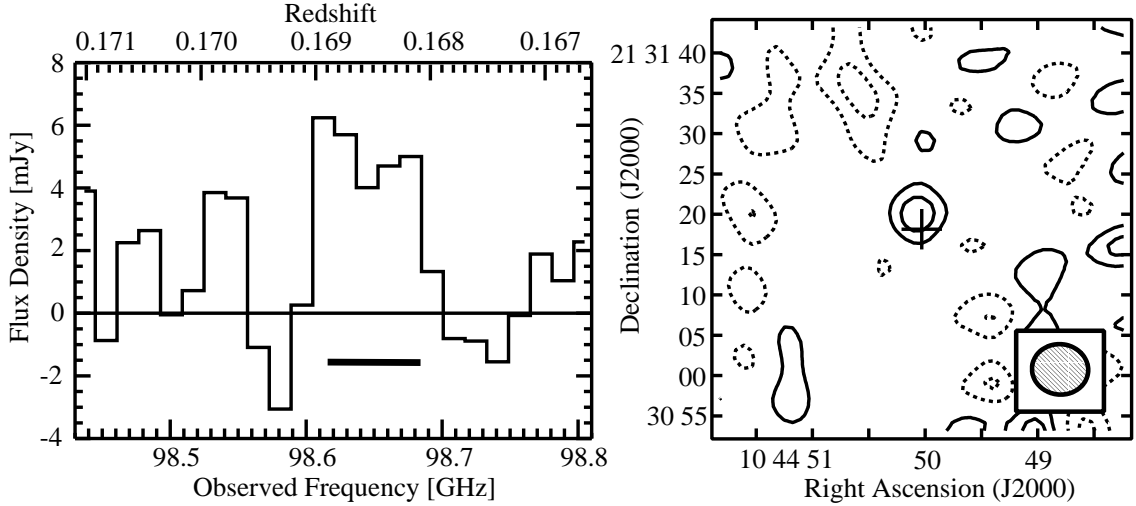}
  \end{center}
\caption{
(Left)
Observed spectrum at the 98 GHz band toward GRB 030329, 
produced by summing up the 3 mm data from 2003 April 10 to 2003 May 30. 
The spectral resolution is 16 MHz or 48.6 km s$^{-1}$, 
and the rms noise level is 3.8 mJy.
A possible emission feature is seen at the redshift of GRB 030329 ($z$ = 0.1685). 
The velocity width of the possible emission feature is $220 \pm 49$ km s$^{-1}$.
A horizontal bar indicates a velocity width of 200 km s$^{-1}$ for reference.
(Right) 
Total intensity map of a possible emission feature, 
integrated over a velocity width of 220 km s$^{-1}$. 
The synthesized beam, shown in the right bottom corner, 
is \timeform{7".0} $\times$ \timeform{6".3} (position angle = $84^\circ$).
The contour interval is 0.52 Jy beam$^{-1}$ km s$^{-1}$ (1 $\sigma$).
The source coincides with the position of the GRB 030329 optical afterglow.
}\label{fig:CO}
\end{figure}

\begin{figure}
  \begin{center}
    \FigureFile(150mm,150mm){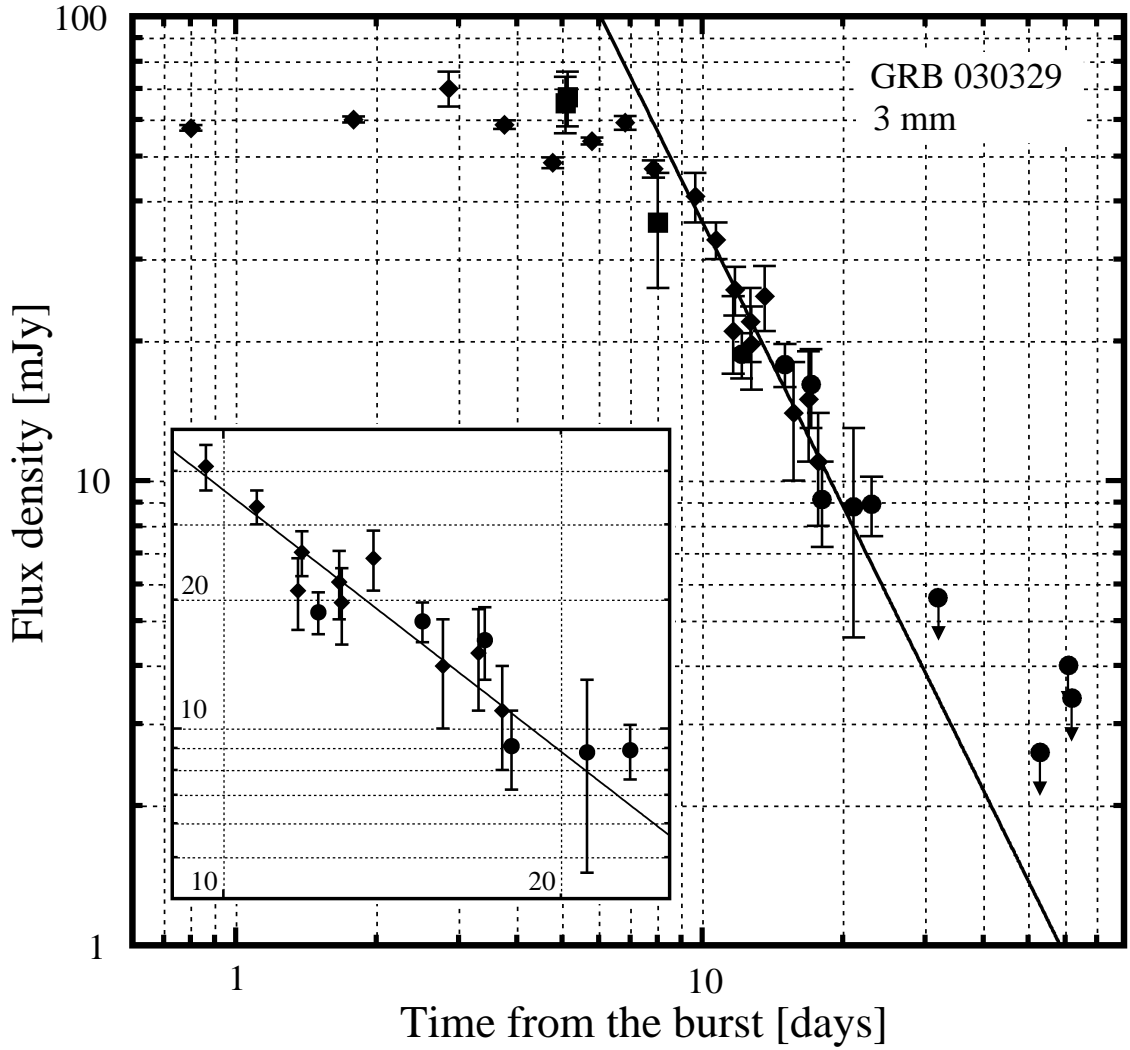}
  \end{center}
\caption{
Millimeter-wave ($\lambda$ = 3 mm) light curve of the afterglow GRB 030329 
taken with the NMA (circle; this work), OVRO/BIMA (diamond; Sheth et al.\ 2003)
and NRO 45 m telescope (square; Kuno et al.\ 2004).
A steep decline of the afterglow 
($S \propto t^{-2.03 \pm 0.21}$, indicated by a solid line), is evident.
Insertion is a close-up view of the light curve for $\Delta t$ = 9 to 30 days.
}\label{fig:lightcurve}
\end{figure}


\begin{table*}
\begin{center}
Table~1.\hspace{4pt}Nobeyama Millimeter Array observations of GRB 030329.\\
\end{center}
\vspace{6pt}
\begin{tabular*}{\textwidth}{@{\hspace{\tabcolsep}
\extracolsep{\fill}}p{6pc}cccccccccc}
\hline\hline\\[-6pt]
UT Date         & Start & Stop  & Duration & $\Delta$t & $f_{\rm obs}$ & Flux  & $\sigma$   & Tsys  & Seeing  \\
                & [hr] & [hr] & [hr]    & [days]     &  [GHz]        & [mJy] & [mJy] & [K]   &         \\
\multicolumn{1}{c}{(1)}  & (2)   & (3)   & (4)      & (5)       & (6)           & (7)   & (8)   & (9)   & (10)    \\[4pt]     
\hline
2003 April 6    & 08.77 & 15.58 & 6.5      & 08.23     & 141           & 41.0 & 3.2 & 160 & Bad \\
2003 April 7    & 10.35 & 16.25 & 5.9      & 09.07     & 141           & 42.5 & 2.4 & 220 &     \\
2003 April 10   & 12.75 & 17.58 & 4.8      & 12.15     & 93            & 18.7 & 2.1 & 210 &     \\
2003 April 13   & 09.50 & 15.92 & 6.4      & 15.05     & 93            & 17.8 & 1.9 & 150 &     \\
2003 April 15   & 11.53 & 16.75 & 5.2      & 17.11     & 93            & 16.1 & 3.1 & 360 &     \\
2003 April 16   & 09.87 & 16.75 & 6.9      & 18.07     & 93            &  9.1 & 1.9 & 220 & Bad \\
2003 April 19   & 11.45 & 16.58 & 5.1      & 21.10     & 93            &  8.8 & 4.2 & 260 & Bad \\
2003 April 21   & 09.42 & 16.58 & 7.2      & 23.06     & 93            &  8.9 & 1.3 & 140 & Good \\
2003 April 23   & 09.52 & 14.92 & 5.4      & 25.03     & 93            &  --- & --- & Bad &     \\
2003 April 30   & 08.43 & 15.50 & 7.1      & 32.02     & 93            &  --- & 2.8 & 200 & Bad \\
2003 May 21     & 08.33 & 14.25 & 5.9      & 52.99     & 93            &  --- & 1.3 & 180 & Good \\
2003 May 22     & 07.10 & 13.92 & 6.8      & 53.95     & 93            &  --- & --- & Bad &     \\
2003 May 29     & 08.43 & 13.58 & 5.2      & 60.98     & 93            &  --- & 2.0 & 210 &     \\
2003 May 30     & 08.50 & 13.42 & 4.9      & 61.97     & 93            &  --- & 1.7 & 200 &     \\
\hline
\end{tabular*}

\vspace{0.5cm}

{\footnotesize
(1) Observing date. (2) Beginning time of observation. (3) End time of observation.
(4) Duration of the observation. Net on-source integration time is typically about half of this.
(5) Time since the burst (2003 March 29.48 UT). Middle of the observing run is taken here.  
(6) Observing frequency. An average of USB and LSB frequencies. 
``141 GHz'' means an average of $146.969 \pm 0.5$ GHz and $134.969 \pm 0.5$ GHz bands, and
``93 GHz'' means an average of $98.824 \pm 0.5$ GHz and $86.824 \pm 0.5$ GHz bands, respectively.
(7) Flux density. (8) Noise level (1 $\sigma$). (9) System noise temperature in DSB.
(10) Atmospheric phase stability. 
``Bad'' means a large fraction of the visibility data should be discarded
due to large phase noise.
}

\end{table*}

\end{document}